\documentclass[3p,onecolumn,times]{elsarticle}
\usepackage[usenames]{color}
\usepackage{graphicx}
\usepackage{amssymb}
\usepackage{amsmath}
\usepackage{epstopdf}
\usepackage[normalem]{ulem}
\journal{Physics Letters A}
\begin{document}
\biboptions{sort&compress}

\begin{frontmatter}
\title{Robust Dynamics of Rogue waves, Breathers and Mixed Bound State Solutions in Spin -Orbit and Rabi Coupled Condensates}

\author[psg]{P.~S.~Vinayagam\corref{cor1}}
\ead{psvinayagam11@gmail.com}
\cortext[cor1]{Corresponding author}
\author[gcw]{R.~Radha}
\ead{vittal.cnls@gmail.com}
\address[psg]{Physics Department, PSG college of Arts and Science (Autonomous),\\ Avinashi road, Civil Aerodrome post, Coimbatore-641 014, Tamil Nadu, India.}
\address[gcw]{Centre for Nonlinear Science (CeNSc), PG and Research Department of Physics, Government College for Women (Autonomous),\\ Kumbakonam-612 001, Tamil Nadu, India.}

\begin{abstract}
In this paper, based on an integrable model governed by four system parameters, namely, spin-orbit coupling and Rabi coupling which are constants while the other two parameters, namely the  harmonic trap and  scattering lengths which are time dependent,   we investigate the spin orbit-Rabi coupled   condensates governed by the two coupled Gross-Pitaevskii(GP) equation. Employing Darboux transformation approach with nontrivial seed solution, we generate rogue waves, breathers, mixed rogue-dark-bright and classical dark-bright solitons. While the addition of spin orbit coupling contributes to rapid oscillations in the amplitude of the rogue waves, Rabi coupling results in the appearance of  stripes in the temporal direction . When the transient trap is switched on, these stripes which remain stable within the confining region, begin to overlap exponentially in the expulsive domain and eventually shrink leading to instability. We have shown  that the instability arising in the  rogue waves in a transient trap can be completely overcome  by manipulating the scattering length through Feshbach resonance. In the case of breathers,the  Rabi coupling introduces temporal stripes   with single and double mode peaks around the origin while under the influence of the transient trap, the breathers get compressed and tilted. On the other hand, the amplitude of breathers which stays constant between a maxima and minima despite oscillating with time undergoes rapid  fluctuations  in a given spatial domain under the influence of SOC. In the case of classical dark-bright solitons, we see the flipping of dark solitons to attain positive density much similar to bright solitons under the impact of Rabi coupling. One also witnesses a 45$^{\circ}$ shift in the  trajectory of dark and bright soltons  when the  transient trap is switched on. The width of the solitons widen in the confining trap while they shrink in the expulsive domain. 
%The transient trap is also found to introduce more oscillations in the real part of the order parameter which is further enhanced by SO coupling.   
\end{abstract} 
\end{frontmatter}

\section{Introduction}
Spin-orbit coupling \cite{r1} through which the the quantum particle's spin interacts with its momentum has triggered a resurgence   in the investigation of condensed matter physics and it has led  to several interesting concepts like spin Hall effect\cite{r2}, topological insulators\cite{r3}, spintronic devices\cite{r4} etc,. The recent experimental realization of spin orbit coupling in ultra cold atoms\cite{r5} has led to a flurry of activities in the domain of Bose Einstein condensates(BECs) to explore several exotic phenomena like supersolids\cite{r6}, spin microemulsion\cite{r7} etc,. Rabi coupling, on the other hand describes the interaction between various spin states through an external electromagnetic field leading to striped solitons \cite{psvcnsns_soc}, coherent oscillations\cite{rabi_coherentos}, phase separation\cite{rabi_phaseseparation} etc,.The dynamics of spin orbit and Rabi coupled BECs is governed by coupled Gross-Pitaevskii(GP) type equation which belongs to the category of variable coefficient coupled nonlinear Schrödinger equations(CNLS). The  surge in the study of coupled nonlinear Schrödinger equations (CNLSEs)  mostly dominated by numerical approaches due to their wide applications in various domains like  nonlinear optics \cite{nlo}, plasma physics\cite{plasma}, bio-physics \cite{biophy} and multi component Bose-Einstein condensates (BECs) \cite{multibec} has acted as a catalyst to penetrate deep into the domain of spin orbit  and Rabi coupled BECs to unearth other unexplored nonlinear excitations.

%Spin orbit coupling which accounts for the intrinsic interaction between a particle's spin and motion is responsible for various important phenomena ,ranging from atomic fine structure to topological condensed matter physics.  Rabi coupling, on the other hand describes the interactions between various spin states by an external electromagnetic field, which significantly impacts the dynamics of the system such as striped solitons \cite{psvcnsns_soc}, coherent oscillations \cite{rabi_coherentos}, and phase separation \cite{rabi_phaseseparation} etc,.

%\textcolor{blue}{The abundant freedom arising by virtue of the combination of SOC, Rabi coupling,  non-linearity and trap frequency in GP type equations or CNLSEs can provide a perfect platform to witness new solitonic structures, instability patterns, its stabilization scenario and complex wave dynamics \cite{socstab}. Understanding these  effects is essential for  controlling and manipulation of spinor wave functions in various experimental settings. Moreover, the theoretical insights gained by  the above investigations  can trigger  novel applications in quantum information processing \cite{infor} , spintronics\cite{spin}, and the design of advanced photonic\cite{photonic} devices.}

It should be pointed out that even though one has come across several investigations in the domain of spin orbit coupled BECs in the last few decades {\cite{lightsoc_r1,super_r2,mi_r3,solid_r4} and localized solutions like bright solitons\cite{sobright}, magnetic solitons\cite{rr1}}, vortices\cite{sovortices}, skyrmions\cite{soskymions} have been identified, 
%\textcolor{blue}{In addition to this, impact of light induced SOC in the phase of density profiles is witnessed in \cite{lightsoc_r1}, Superstripes and the excitation spectrum using Bogoliubov theory is addressed in \cite{super_r2}, Modulational instability in 1d BEC is discussed in \cite{mi_r3}, dynamics of stripes pattern in supersolid SOC bose gases is investigated in \cite{solid_r4}, and witnessed magnetic solitons numerically in \cite{rr1}}. 
nonlinear excitations like  rogue waves, breathers, mixed bound states etc,. and how their dynamics are influenced by either SOC or Rabi coupling have not yet been reported analytically.  
%\textcolor{blue}{ The fact that the nonlinear excitations do not superpose and  highly unpredictable besides being mathematically complex unlike the linear counterparts underlines the significance of this investigation}
It should also be emphasized that the dynamics of spin orbit-Rabi coupled BECs has not yet been analytically solved  from the perspective of integrability  as well and most of the investigations have been dominated only by numerical simulations.  Motivated by the above consideration, we have explored the dynamics of spin orbit-Rabi coupled BECs  based on an integrable model involving four system parameters with two of them, namely SOC and Rabi coupling being constants and the other two, namely trap frequency and scattering length being time dependent in an attempt to generate other nonlinear excitations. The fact that the nonlinear excitations do not superpose and  highly unpredictable besides being mathematically complex unlike the linear counterparts underlines the significance of this investigation. In particular, we wish to generate  rogue waves  which are highly unstable  appearing  from nowhere and vanishing  without a trace \cite{roguebasic}  and breathers which oscillate with time analytically  \cite{breatherbasic}. In addition, an attempt to stabilize rogue waves using Feshbach resonance is also being explored. 

The paper is organized as follows. In this paper, we plan to  investigate the dynamics of spin orbit and Rabi coupled BECs described by coupled Gross-Pitaevskii(GP) type equation. Section II describes the mathematical model and the Lax pair of the problem under investigation. Section III describes the general scheme of Darboux transformation method. In section IV,  we employ Darboux transformation approach and construct nonlinear excitations like rogue waves, breathers, bright and dark soliton solutions starting from a nontrivial seed solution. We then study the interplay between spin orbit coupling, Rabi coupling, time dependent trap strength and time dependent scattering parameter on the nonlinear excitations. Finally, we conclude with the results of the investigation in section V. 

\section{Model and Lax pair}
Considering a spin orbit  coupled quasi one dimensional BEC in a  harmonic  trap with longitudinal and transverse frequencies satisfying the relation $\omega_x \ll \omega_\perp$, it can be described at sufficiently low temperatures by the  two coupled GP equation of the following form

\begin{subequations}\label{eq:gpe}
\begin{align}
\mathrm{i} \partial_t q_{1}
   = & \left[-\partial_x^2+V_{\text{trap}}(x,t)-2\gamma(t)(\left\vert q_{1}\right\vert^2+\left\vert q_{2}\right\vert^2) +\mathrm{i}k_L\partial_x \right] q_{1}
        -\Omega q_{2}\label{eq:gpe1a}, \\
\mathrm{i} \partial_t q_{2}
   = & \left[-\partial_x^2+V_{\text{trap}}(x,t)-2\gamma(t)(\left\vert q_{1}\right\vert^2+\left\vert q_{2}\right\vert^2) -\mathrm{i}k_L\partial_x \right] q_{2}
       -\Omega q_{1} \label{eq:gpe1b}.
\end{align}
\end{subequations}%
In the above equation, $q_{i}$, i=1,2 are field variables, $\pm i k_L \partial_x$ represents the momentum transfer between the laser beams and the atoms due to SOC,
$\gamma(t)=\gamma_{0}\; exp \int \sigma(t) dt$ represents binary  interaction(scattering length), which depends on the choice of $\sigma(t)$ defined by Eq. \eqref{sigmaoneta} as $\frac{d}{dt}log(\eta(t))$, where $\eta(t)$ is an arbitrary time dependent parameter \cite{serkin}. $\Omega$ denotes Rabi coupling and $V_{\text{trap}}(x,t)=(\lambda(t))^2 x^2 /2$, where $\lambda(t)={\omega_x}/{\omega_\perp}$. To construct  soliton solutions, we begin with  the celebrated Manakov model. It should be mentioned that  the model equations (\ref{eq:gpe1a}) and (\ref{eq:gpe1b}) reduce to the celebrated Manakov model by eliminating SO coupling and Rabi coupling one by one in  the following manner. By considering  $V_{\text{trap}}(x,t)$=0 and  employing the following transformation
\begin{subequations}
\begin{align}
    q_1(x,t) &= q_1(x,t)\; exp \Bigg[\frac{i}{2}k_{L} t - 2 x\Bigg] \\
    q_2(x,t) &= q_2(x,t)\; exp \Bigg[\frac{i}{2}k_{L} t + 2 x\Bigg]
\end{align}\label{soremove}
\end{subequations}
we convert the two coupled GP equation   (\eqref{eq:gpe}) into constant coeffienct NLS equation (\eqref{eq:gpewithrabi}) after elimination of SO coupling as 
\begin{subequations}\label{eq:gpewithrabi}
\begin{align}
\mathrm{i} \partial_t q_{1}
   = & \left[-\partial_x^2-2(\left\vert q_{1}\right\vert^2+\left\vert q_{2}\right\vert^2)\right] q_{1}
        -\Omega q_{2}\label{eq:gpe1awithrabi} \\
\mathrm{i} \partial_t q_{2}
   = & \left[-\partial_x^2-2(\left\vert q_{1}\right\vert^2+\left\vert q_{2}\right\vert^2) \right] q_{2}
       -\Omega q_{1} \label{eq:gpe1bwithrabi}
\end{align}
\end{subequations}%
By invoking the following transformation, 

\begin{align}\label{treq}
\left(
\begin{array}{c}
    q_1(x,t) \\
    q_2(x,t) \\
\end{array}
\right)=\left(
	\begin{array}{cc}
		a \cos (\Omega  t) & b \sin (\Omega  t) \\
		b \sin (\Omega  t) & a \cos (\Omega  t) \\
	\end{array}
	\right).\left(
	\begin{array}{c}
		\Tilde{q_1}(x,t) \\
		\Tilde{q_2}(x,t) \\
        \end{array}
	\right).
\end{align}
where  the constants are  $a=1$  $\&$  $b=-i$, we can remove Rabi coupling  to convert eq.(3) to the celebrated Manakov model (after dropping the bar)
\begin{subequations}\label{eq:manakov}
\begin{align}
\mathrm{i} \partial_t q_{1}
   = & \left[-\partial_x^2-2(\left\vert q_{1}\right\vert^2+\left\vert q_{2}\right\vert^2)\right] q_{1}
       \label{eq:manakov1a} \\
\mathrm{i} \partial_t q_{2}
   = & \left[-\partial_x^2-2(\left\vert q_{1}\right\vert^2+\left\vert q_{2}\right\vert^2) \right] q_{2}
       \label{eq:manakov1b}
\end{align}
\end{subequations}%
The above Manakov model admits  the following Lax pair \cite{psvrotation} 
\begin{subequations}\label{laxcondition}
	\begin{align}
		{\bf \Phi}_{x}& ={\bf U} {\bf \Phi} = {\bf U}_{0}\,{\bf \Phi} + {\bf U}_{1}\,{\bf
			\Phi}\,{\bf \Lambda}\,,  \\
		{\bf \Phi}_{t}& = {\bf V} {\bf \Phi} = {\bf V}_{0}\,{\bf \Phi} + {\bf V}_{1}\,{\bf
			\Phi}\,{\bf \Lambda} + {\bf V}_{2}\,{\bf \Phi}\,{\bf \Lambda}^{2}\,,
	\end{align}
\end{subequations}
where,
%\begin{widetext}
\begin{align}
	{\bf U}_{0}=
	\begin{pmatrix}
		0 & q_{1}(x,t) & q_{2}(x,t) \\\\
		-q_{1}^{\ast}(x,t) & 0 & 0 \\\\
		-q_{2}^{\ast}(x,t) & 0 & 0%
	\end{pmatrix},
	\;\;\; {\bf U}_{1}=
	\begin{pmatrix}
		1 & 0 & 0 \\\\
		0 & -1 & 0 \\\\
		0 & 0 & -1%
	\end{pmatrix},
 \;\;\; {\bf \Lambda}& =
	\begin{pmatrix}
		\zeta_1 & 0 & 0 \\\\
		0 & \zeta_2 & 0 \\\\
		0 & 0 & \zeta_3
	\end{pmatrix},\notag
	%\label{U}
\end{align}
%\end{widetext}
%\begin{widetext}
\begin{align}
	{\bf V}_{0}=\frac{\mathrm{i}}{2}
	\begin{pmatrix}
		q_{1}(x,t) q_{1}^{\ast}(x,t)+ q_{2}(x,t) q_{2}^{\ast}(x,t)  & q_{1x}(x,t) & q_{2x}(x,t)
		\\\\
		q_{1x}^{\ast}(x,t) & -q_{1}(x,t) q_{1}^{\ast}(x,t) & -q_{2}(x,t) q_{1}^{\ast}(x,t) \\\\
		q_{2x}^{\ast}(x,t) & -q_{1}(x,t) q_{2}^{\ast}(x,t) & -q_{2}(x,t) q_{2}^{\ast}(x,t)%
	\end{pmatrix}, \notag
\end{align}
%\end{widetext}
%\begin{widetext}
\begin{align}
	{\bf V}_{1} &=
	\begin{pmatrix}
		0 & -q_{1}(x,t) & -q_{2}(x,t) \\\\
		q_{1}^{\ast}(x,t) & 0 & 0 \\\\
		q_{2}^{\ast}(x,t) & 0 & 0%
	\end{pmatrix},
	\;\;\; {\bf V}_{2}=\mathrm{i}
	\begin{pmatrix}
		1 & 0 & 0 \\\\
		0 & -1 & 0 \\\\
		0 & 0 & -1%
	\end{pmatrix}, \notag
	%\label{V}
\end{align}\\
where $\zeta_{1,2,3}$ are  the spectral parameters. $\Phi =(\phi _{1},\phi_{2},\phi _{3})^{T}$ is a three-component Jost function, ${\bf U}$ and ${\bf V}$, known as the Lax pair, are functionals of the
solutions of the model equations.The consistency condition ${\bf \Phi}_{xt}={\bf \Phi}_{tx}$ leads to $ {\bf U}_{t}-{\bf V}_{x}+[{\bf U},{\bf V}]={\bf 0}$, which is equivalent to  the Coupled Nonlinear Scr\"odinger equations (CNLSE) termed as Manakov model.
Mapping of Manakov to coupled Gross-Pitaevskii equation can  be done by employing  similarity transformation given in \cite{psvpwp}. While invoking the  similarity transformation, the following constraints will have to be complied with  for the successful mapping of Manakov to coupled GPE. They are of the following form
\begin{align}\label{sigmaoneta}
    \frac{d}{dt}\eta(t)=\sigma(t) \eta(t)
\end{align}
 with
 \begin{align}\label{tr}
    \lambda(t)^2=\frac{d}{dt}\sigma(t)-\sigma(t)^2
\end{align}
and
\begin{align}\label{sl}
    \sigma(t)=\frac{d}{dt} ln \gamma(t) \
\end{align}
From equation \eqref{sigmaoneta}, one can choose  $\sigma(t)$ satisfying the condition $\sigma(t)=\frac{d}{dt}log(\eta(t))$, where, $\eta(t)$ is an arbitrary time dependent parameter. Accordingly,  one can fix the nature of the trapping potential $\lambda(t)$  given by \eqref{tr} and scattering length $\gamma(t)$  by equation \eqref{sl}. Equation \eqref{tr} represents the parabolic(harmonic) trapping potential $\lambda(t)^2$, related to the interaction strength $\gamma(t)$  of the coupled GP equation through the integrability condition  given in \cite{psvcnsns_soc,serkin}.

\section{Darboux transformation}
It is well known that the Darboux transformation for eq. \eqref{eq:gpe} is given by \cite{matveev}
\begin{align}
    T &= \zeta I - [\zeta_1^{\ast}- (\zeta_1^{\ast}-\zeta_1) P_1],\;\;   P_1 = \frac{\phi_1 \phi_1^{\dag}}{\phi_1^{\dag} \phi_1}
\end{align}
where  $\phi_1$=$\phi(x,t,\zeta_1)$$(m_1,m_2,m_3)^{\text{Tr}}$, $m_1,m_2,m_3$ are constants, $P_1$ is a projection matrix, $I$ is 3 $\times$ 3 identity matrix and $\phi(x,t,\zeta_1)$ is the fundamental solution of the Lax pair equation at $\zeta=\zeta_1$, $q_i=q_i[0]$($i$=1,2) which leads to the following transformation between the fields
\begin{subequations}\label{dttemplate}
    \begin{align}
    q_1[1] &= q_1[0]+ 2 I \frac{(\zeta_1-\zeta_1^{\ast}) \phi_1 \phi_2^{\ast}}{\lvert \phi_1 \rvert^{2} + \lvert \phi_2 \rvert^{2} + \lvert \phi_3 \rvert^{2}}, \\
    q_2[1] &= q_2[0]+ 2 I \frac{(\zeta_1-\zeta_1^{\ast}) \phi_1 \phi_3^{\ast}}{\lvert \phi_1 \rvert^{2} + \lvert \phi_2 \rvert^{2} + \lvert \phi_3 \rvert^{2}}
\end{align}
\end{subequations}
In the above equation, $q_i[j]$ represents the field variables where the $j=0,1,2$ indicates its zero order, first order and second order iteration. In other words, $q[0]$ represents the seed (vacuum) solution and $q[1]$ the first iterated solution and so on. 
%%%%%fig1
\section{Rogue waves, Breathers, Bright and Dark solitons}
It should be emphasized that  seed solutions play a significant role by helping us obtain the required form of localized solutions such as rogue waves, breathers and classical solitons like bright or dark solitons.  Hence, one can obtain the required form of localized excitations in Darboux transformation by a careful selection of the seed solutions.   Rogue waves are inherently transient in nature and localized in both space and time while breathers exhibit periodicity and stable within their oscillatory domains. This distinction stems from the underlying symmetry and the chosen eigenvalues during the process of Darboux transformation. Since rogue waves, breathers and dark solitons are intrinsically related to modulated wave backgrounds, one has to choose a non zero seed solution. The physics of these phenomena necessitates the existence of a complex interplay  between their background wave and localized structures. Hence, for rogue waves, breathers and dark solitons, one has to begin with non zero plane wave as the seed solution. Since bright solitons are self contained solutions of the models arising  due to the delicate balance between the dispersion and nonlinearity, they are exempted from non zero seed. They do not rely on background field :Instead, they are stand alone structure where, all the energy is localized within the soliton itself. The selection of nonzero plane wave as the seed solution for both field variables leads to rogue waves whereas  selection of nonzero plane wave for one field and zero seed for another field variable leads to breathers. 
\subsection{Rogue waves}
\begin{figure}
\centering
\includegraphics[scale=0.40]{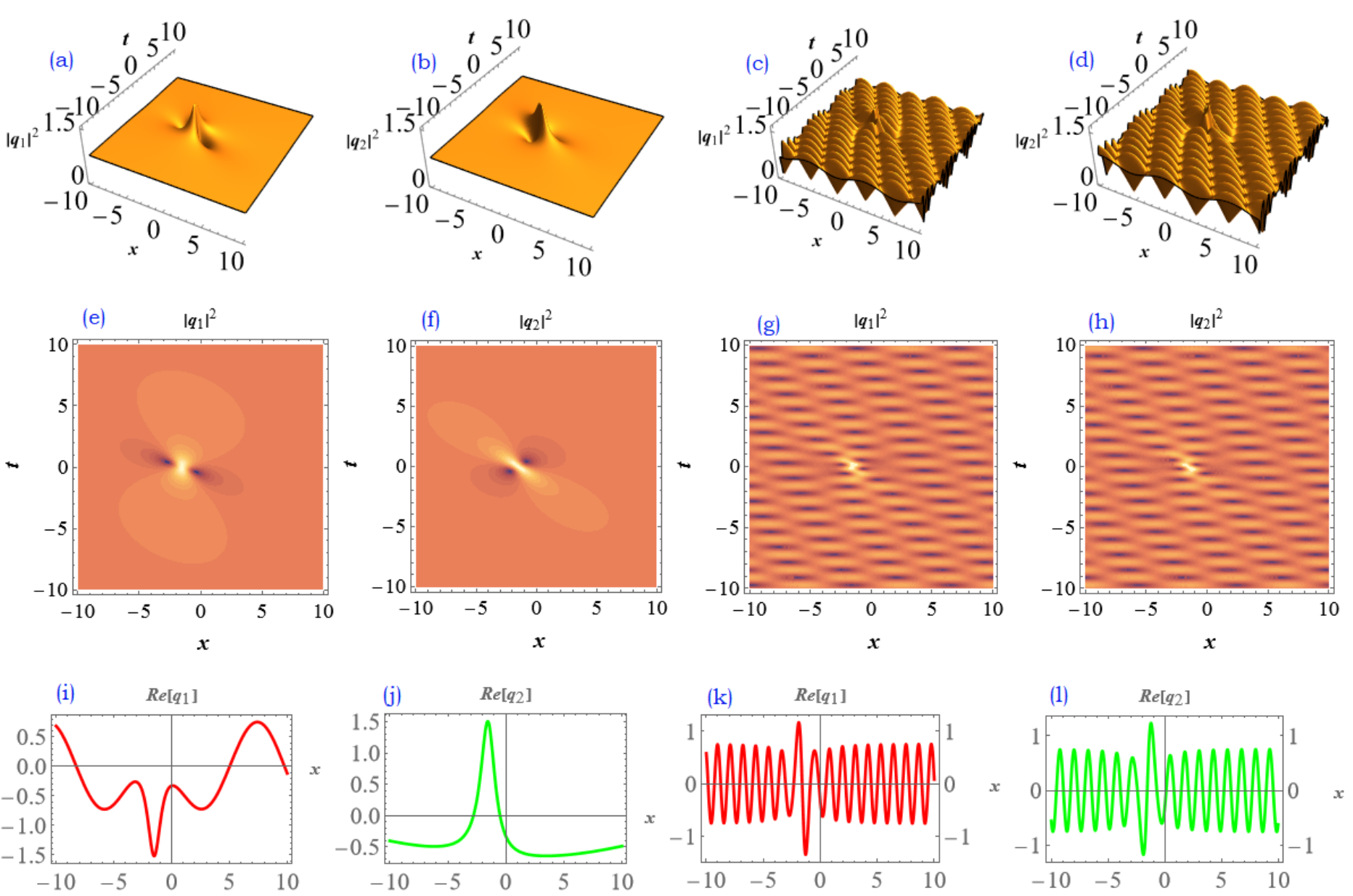}
\caption{First-order rogue wave without trap: For the parametric choice $\sigma(t)=0$, $\alpha=d_2+3 e_1$, $c_{1}=c_{2}=2\alpha$, $d_1=d_2=2\alpha$, and $d_2=0.08$, $e_1=0.1$. \textbf{First row:} (a, b) Rogue waves without Rabi coupling $(\Omega=0)$. \textbf{(c, d)} Rogue waves with Rabi coupling $( \Omega=2)$. \textbf{Middle row:} (e, f) and (g, h) represent the corresponding contour plots of the 3D plots in the first row . \textbf{Last row:} (i, j) Real part of the field variables $q_1$ and $q_2$ without SO coupling $(k_L=0)$. \textbf{(k, l)} Real part of the field variables $q_1$ and $q_2$ with SO coupling $(k_L=8)$.}
\label{Figoner1}
\end{figure}

In order to derive the rational solutions,we employ Darboux transformation method and 
choose the seed solution of the following form 
\begin{align}
    q_1[0] &= c_1 e^{i \theta_1(x,t)}, \\
    q_2[0] &= c_2 e^{i \theta_2(x,t)}
\end{align}
where $\theta_{i}(x,t)= d_i x + (2 c_1^{2} + 2 c_2^{2} - d_i^2) t$ with $c_i$ and $d_i$  being  arbitrary constants. We now insert  these seed solutions into the Lax pair equations \eqref{laxcondition} and transform it into the following form
\begin{align}
    \Psi_x =& (MUM^{-1} + M_x M^{-1}) \Psi = U_1 \Psi, \\
    \Psi_t =& (MUM^{-1} + M_x M^{-1}) \Psi = V_1 \Psi
\end{align}
where
$\Psi=M \phi, $ $M= diag \{exp(-(\theta_1(x,t)+\theta_2(x,t)),exp(2\theta_1(x,t)-\theta_2(x,t)),exp(-(2\theta_2(x,t)-\theta_1(x,t))\}$. Accordingly, $U_1$ and $V_1$ can be computed as
\begin{align}
    {\bf U}_{1}=
	\begin{pmatrix}
		-2 i \zeta_1-i(d_1+d_2) & c_1 & c_2 \\\\
		-c_1 & i \zeta_1+i(2d_1-d_2) & 0 \\\\
		-c_2 & 0 & i \zeta_1+i(2d_2-d_1) %
	\end{pmatrix}, \notag
\end{align}
and $V_1 = i U_1^2-\frac{2}{3}(d_1+d_2)-2\zeta_1)U_1+m I$ where, $m= 2i \zeta_1^2+\frac{2}{3} i (c_1^2 + c_2^2)+\frac{2}{9}(d_1^2-d_1d_2+d_2^2)+\frac{2}{3}i\zeta_1 (d_1+d_2)$. 
We choose the parameters to satisfy the relation $c_1=c_2=\pm 2\alpha, d_1=d2-2\alpha, \alpha=d_2 + 3 e_1, e_1= Re(\zeta_1), Im(\zeta_1)=\pm\sqrt{3\alpha}$ where  $d_1$ and  $Re(\zeta_1)$ are arbitrary real numbers and after tedious calculation, we obtain the fundamental solution for the lax pair matrix at $\zeta=\zeta_1$, $q_i=q_i[0]$.  By choosing the parameters $m_1=0, m_2=1, m_3=0$, we arrive at the rogue wave solution given by 
\begin{subequations}\label{rogue}
\begin{align}
	q_{1}(x,t)&=\alpha\;  e^{i \theta_1(x,t)} \left[-1-i \sqrt{3}+\frac{-6 \sqrt{3} \alpha  \delta +i \left(6 \alpha  \delta +5 \sqrt{3}\right)-3}{12
		\alpha ^2 \delta ^2+8 \sqrt{3} \alpha  \delta +5}\right] \\
		q_{2}(x,t)&=\alpha\;  e^{i \theta_2(x,t)} \left[-1+i \sqrt{3}+\frac{-6 \sqrt{3} \alpha  \delta +i \left(-6 \alpha  \delta -5 \sqrt{3}\right)-3}{12
			\alpha ^2 \delta ^2+8 \sqrt{3} \alpha  \delta +5}\right]
\end{align}
\end{subequations}
where $\theta_i(x,t) = d_i x + \left(2 c_1^2+2 c_2^2-d_i^2\right) t$ and  $\delta= x+ 6 e_1 t$ with $c_i$ and $d_i$ being arbitrary constants. For a suitable choice of parameters,  the first order rogue wave is displayed  in figure. \ref{Figoner1}.
\begin{figure}[h]
        \centering
        \includegraphics[scale=0.50]{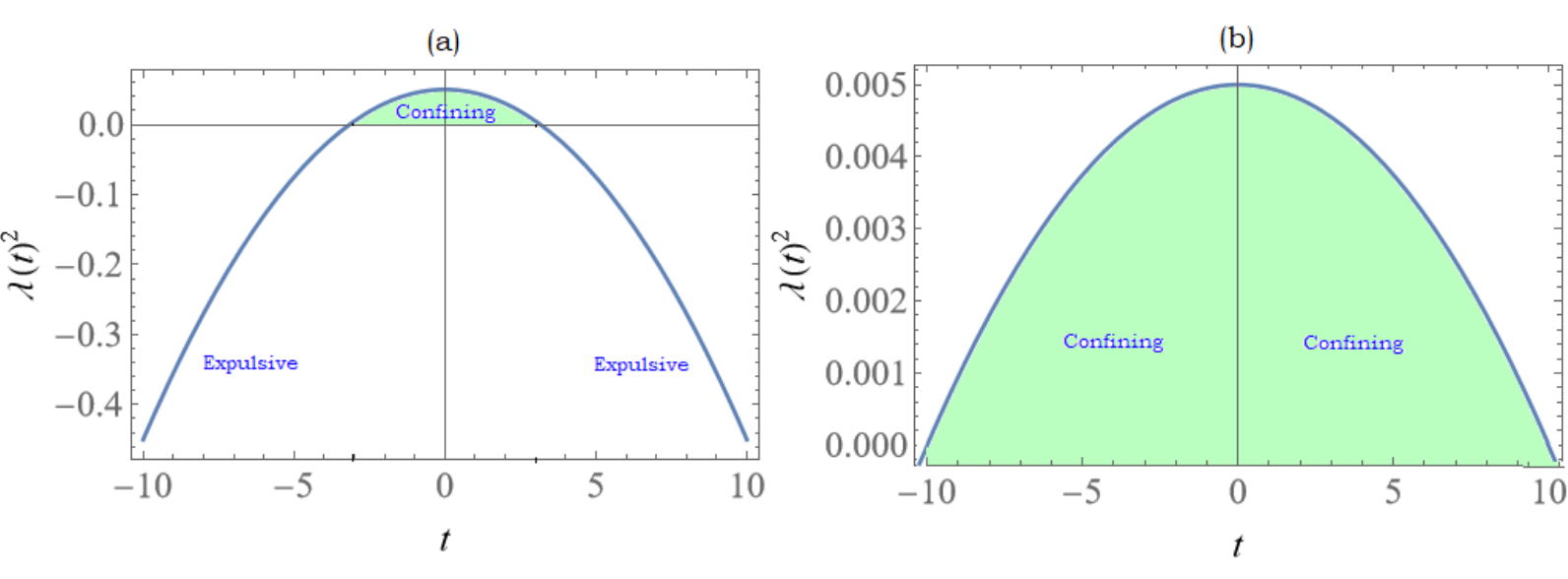}
	\caption{Evolution of  time dependent trapping potential for the choice of parameters (a) $\sigma(t)$= 0.05t and (b) $\sigma(t)$=0.025 t }\label{trap}
\end{figure}
\begin{figure}
\centering
\includegraphics[scale=0.40]{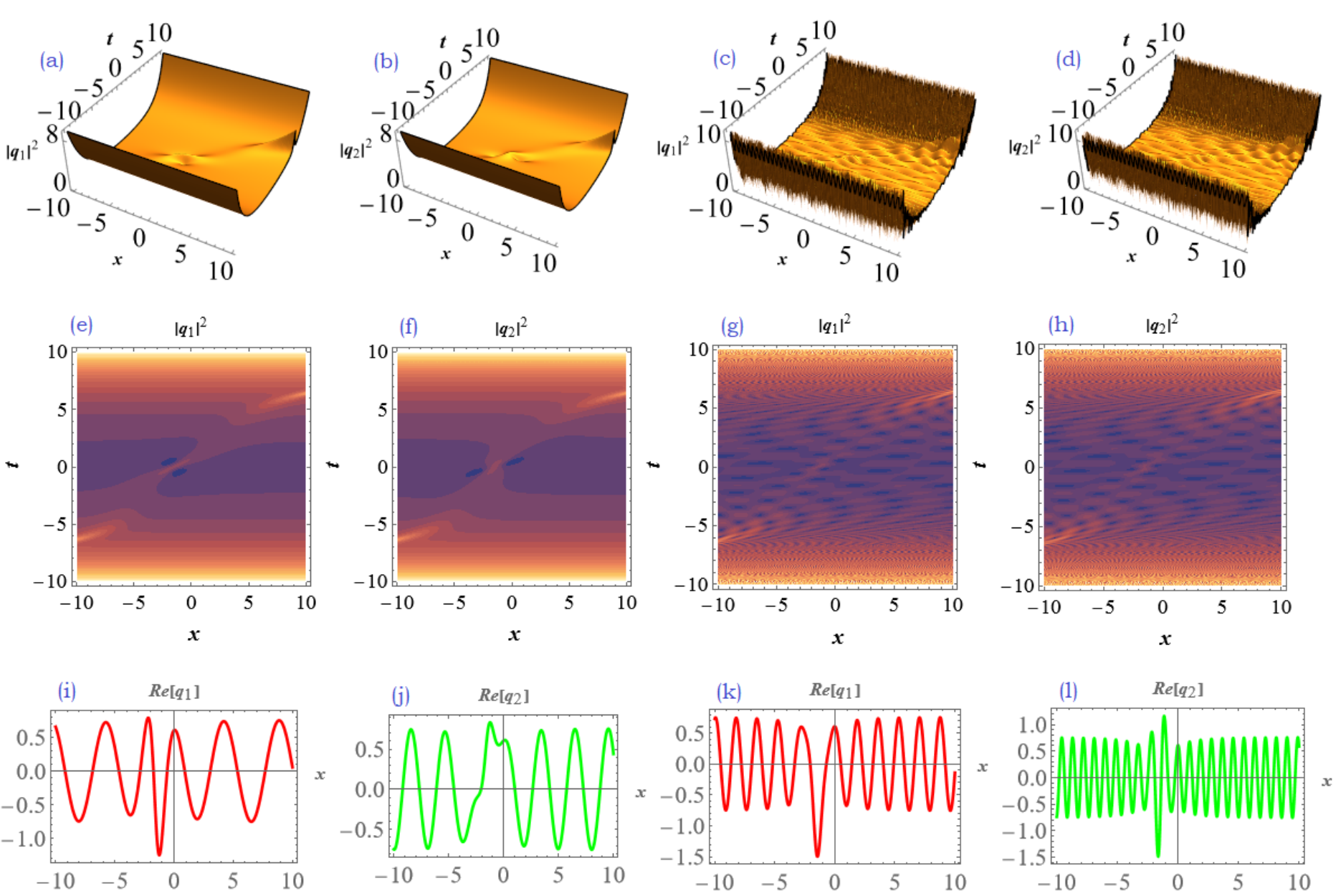}
\caption{First-order rogue wave with transient trap: The rogue waves for the parametric choice $\sigma(t) = 0.05t$, with all other parameters being the same as in Figure~\ref{Figoner1}. \textbf{First row:} (a, b) Rogue waves without Rabi coupling $(\Omega = 0)$. \textbf{(c, d)} Rogue waves with Rabi frequency $(\Omega = 2)$. \textbf{Middle row:} (e, f) and (g, h) are the corresponding contour plots of the 3D figures in the first row. \textbf{Last row:} (i, j) and (k, l) show the real part of the field variables $q_1$ and $q_2$, without SO coupling $(k_L = 0)$ and with SO coupling $(k_L = 8)$, respectively.}
\label{Figoner2}
\end{figure}

\begin{figure}
\centering
\includegraphics[scale=0.40]{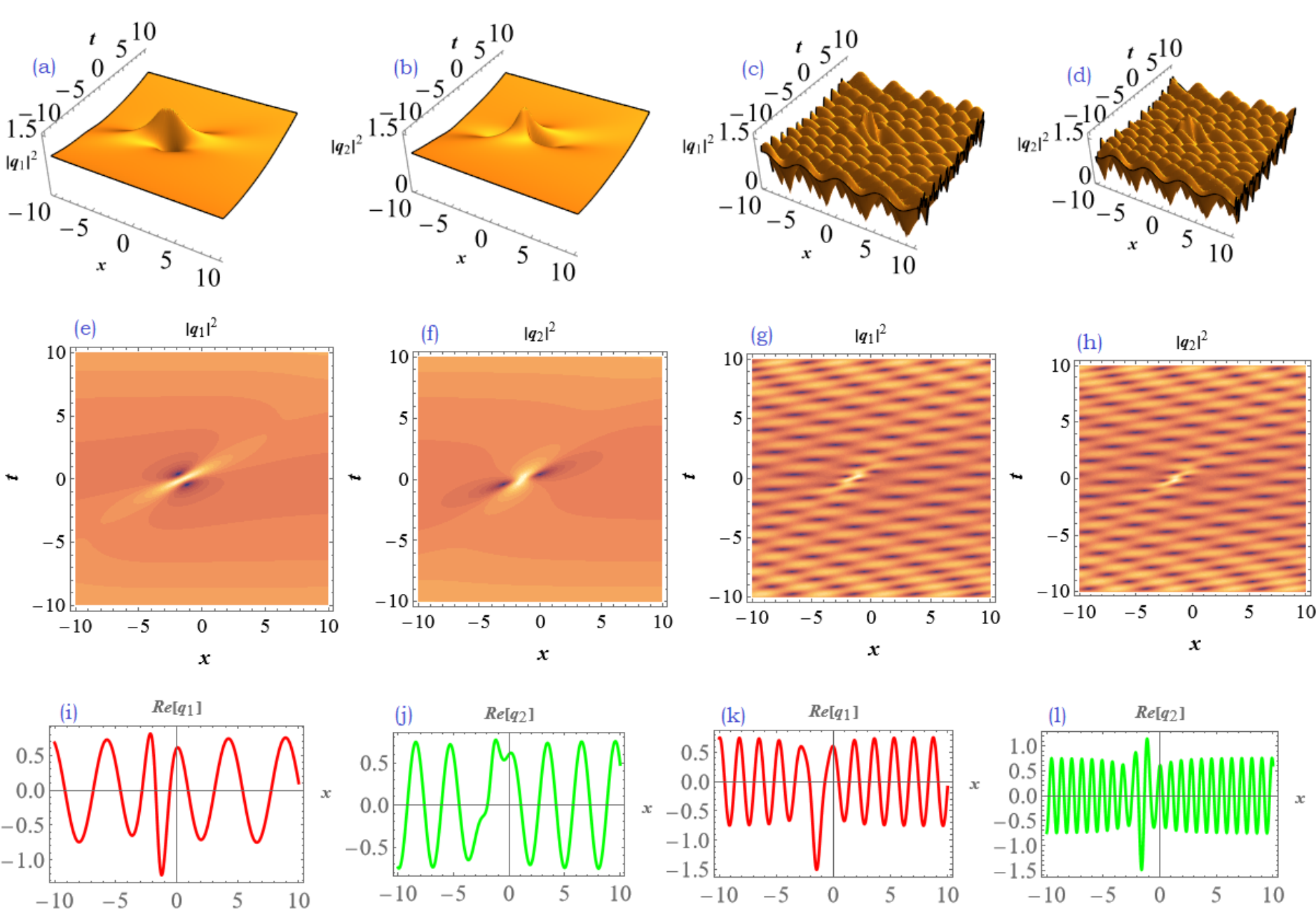}
\caption{Stabilized first-order rogue wave with transient trap: The rogue waves are analyzed for the parametric choice $\sigma(t) = 0.025t$, with all other parameters being the same  as in Figure~\ref{Figoner1}. \textbf{First row:} (a, b) Rogue waves without Rabi coupling $(\Omega = 0)$. \textbf{(c, d)} Rogue waves with Rabi coupling   ($\Omega = 2$). \textbf{Middle row:} (e, f) and (g, h) show the corresponding contour plots of the 3D plots in the first row. \textbf{Last row:} (i, j) and (k, l) depict the real parts of the field variables $q_1$ and $q_2$ without SO coupling $(k_L = 0)$ and with SO coupling $(k_L = 8)$, respectively.}
\label{Figoner3}
\end{figure}

\subsection{Rogue wave without trap} 
The first-order rogue waves described by Eq.~\eqref{rogue}, for the model given by Eq.~\eqref{eq:gpe} without a trapping potential are depicted in Figure~\ref{Figoner1}. The panels in the first row, (a) and (b), illustrate rogue waves in the absence of Rabi coupling ($\Omega = 0$) while  the panels (c) and (d) show the same rogue waves with the inclusion of Rabi coupling ($\Omega=2$). By comparing panels (c, d) with (a, b), it can be observed that the inclusion of Rabi coupling results in the appearance of stripe-like structures along the temporal axis.

In the middle row, panels (e), (f), (g) and (h) represent  the corresponding contour plots of the figures shown in  the first row represented by the panels  (a),( b), (c), and (d) respectively. Panels (e) and (f) show petal like structures of rogue waves while the panels shown in (g) and (h)  reconfirm the appearance of stripes along the temporal axis.  Finally, in the last row, panels (i) and (j) display the real part of the field variables $q_1$ and $q_2$ without spin-orbit (SO) coupling ($k_L=0$) while panels (k, l) show the  real part of the same field variables with SO coupling for  $k_L=8$.  A simple comparison of the panels shown in the last row indicates that the inclusion of SO coupling introduces  rapid oscillations in the real parts of the field variables.

\subsection{Rogue wave with transient trap} 
When the transient trap shown in Fig.~\ref{trap} (a), is switched on, the corresponding first-order rogue wave profiles are depicted in Fig.~\ref{Figoner2}. Panels (a, b) display the rogue waves without the Rabi coupling. Introduction of Rabi coupling ($\Omega = 2$) in the presence of the transient trap results in striped bands along the temporal direction. Notably, these bands appear stable within the confining region and begin to overlap exponentially with the band gaps shrinking in the expulsive region, as shown in panels (c) and  (d).

In the middle row, panels (e), (f), (g) and  (h) show the corresponding contour plots of the rogue waves without and with Rabi coupling shown by the panels (a),( b), (c) and  (d) respectively. Panels (e) and (f) indicate that the petal structure gets disrupted in the transient trap.  The panels (i), (j)  shown in the last row in Fig.~\ref{Figoner2} depict the real part of the field variables $q_1$ and $q_2$ without SO coupling while the panels (k) and  (l) display the real part of the same field variables with SO coupling. It is again evident that SO coupling introduces rapid oscillations in the field variables. In addition, the introduction of the transient trap itself induces  more oscillations, as evident  by comparing the panels (i,j) of  Figs.~\ref{Figoner2}  with that of Figs.~\ref{Figoner1}.

\subsection{Stabilization of rogue waves} 
We observe that the moment the transient trap is switched on, the stripe bands overlap in the expulsive region leading to   instability in the system. To overcome this instability arising in the  rogue wave structures, we manipulate  the scattering length through Feshbach resonance by changing  $\text{Exp}[0.05t]$ to $\text{Exp}[0.025t]$ to  extend the range of the confining region from -10 to 10, as shown in Fig.~\ref{trap} (b) consistent with the integrability constraints given by eqs(7-9). This extended domain of the confining region is in stark contrast to the domain shown earlier in Fig.~\ref{trap} (a) spanning over -3 to 3 along the temporal axis. It should be emphasized that the stabilization of the rogue waves \cite{tame} which are inherently unstable has become possible as we allow the condensates to occupy  the confining region.

The results are presented in Fig.~\ref{Figoner3}.  The first row displays the  rogue waves without Rabi coupling (shown by panels  (a) and (b)) and with Rabi coupling (shown by panels(c) and (d)) while  the corresponding contour plots for the first row are shown in the middle row  by panels (e),(f), (g) and (h) respectively. It is obvious from the panels shown in panels (e) and (f) that we are able to retrieve the  petal structure  shown earlier in panels (e) and (f) of  Fig.~\ref{Figoner1}.  In addition,  we are also able to recover the striped pattern shown in panels (g) and (h) of Fig.~\ref{Figoner1}. The last row shown by panels (i) and (j) show the real part of the field variables $q_1$ and $q_2$ without SO coupling $(k_l=0)$ while  the panels (k) and (l) with SO coupling $(k_l=8)$.

\subsection{Breathers}
\begin{figure}
\centering
\includegraphics[scale=0.40]{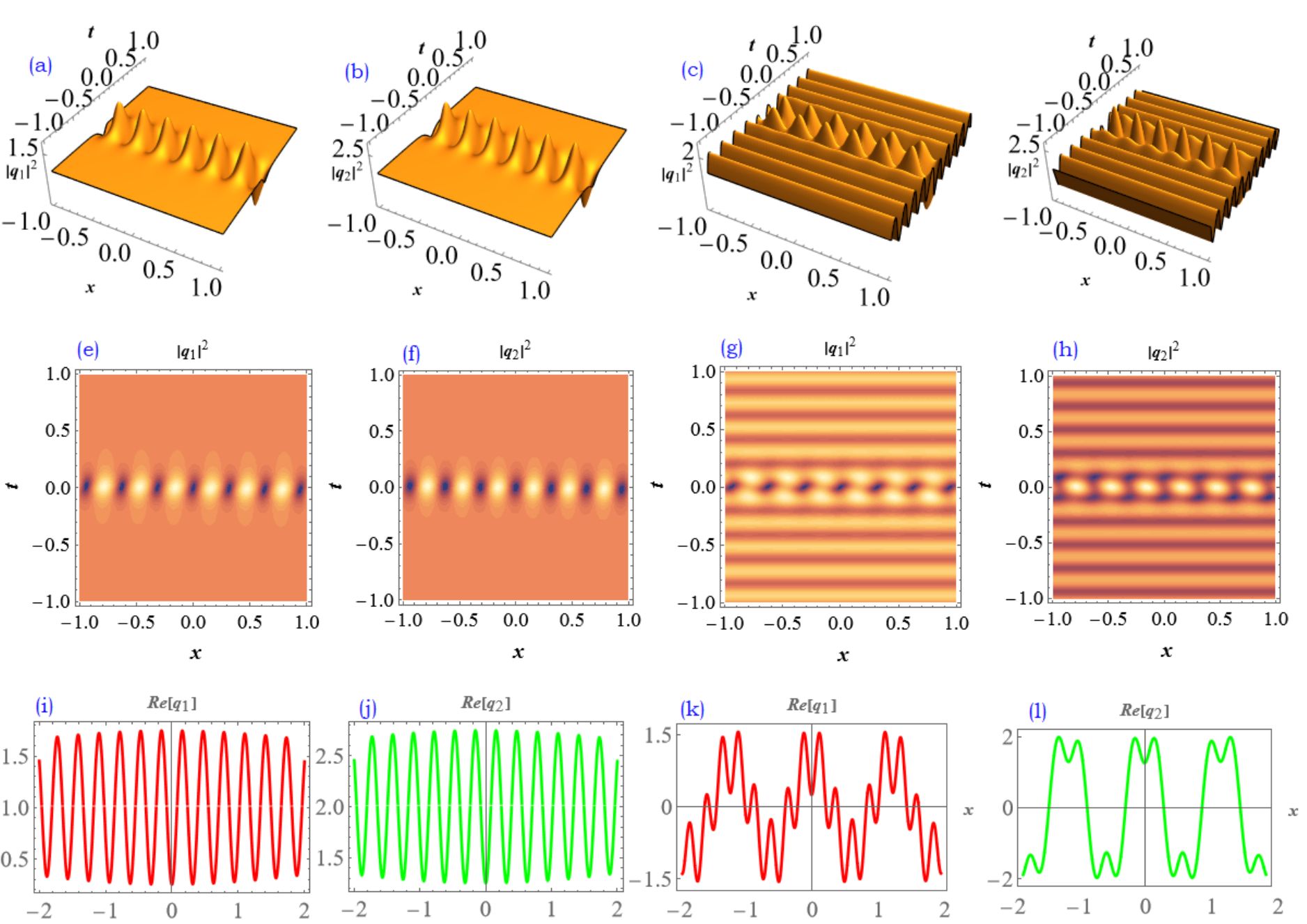}
\caption{\textbf{Breathers Without Trap} The parameters used are $\sigma(t) = 0$, $a_i = b_i = 1$, $c_1 = 1$, $c_2 = 2$, $m_1 = 1$, $\zeta_1 = 10i$, and $\zeta_2 = -0.25i$.\textbf{First row:} (a, b) Breathers without Rabi coupling ($\Omega = 0$); (c, d) Breathers with Rabi coupling ($\Omega = 2$).
\textbf{Middle row:} (e, f) and (g, h) represent the corresponding contour plots corresponding of  the first row  shown by panels (a, b) and (c, d). \textbf{Last row:} (i, j) Real part of the field variables $q_1$ and $q_2$ without SO coupling ($k_L = 0$); (k, l) Real part of the field variables $q_1$ and $q_2$ with SO coupling ($k_L = 8$).}
\label{Figonebr1}
\end{figure}

\begin{figure}
\centering
\includegraphics[scale=0.40]{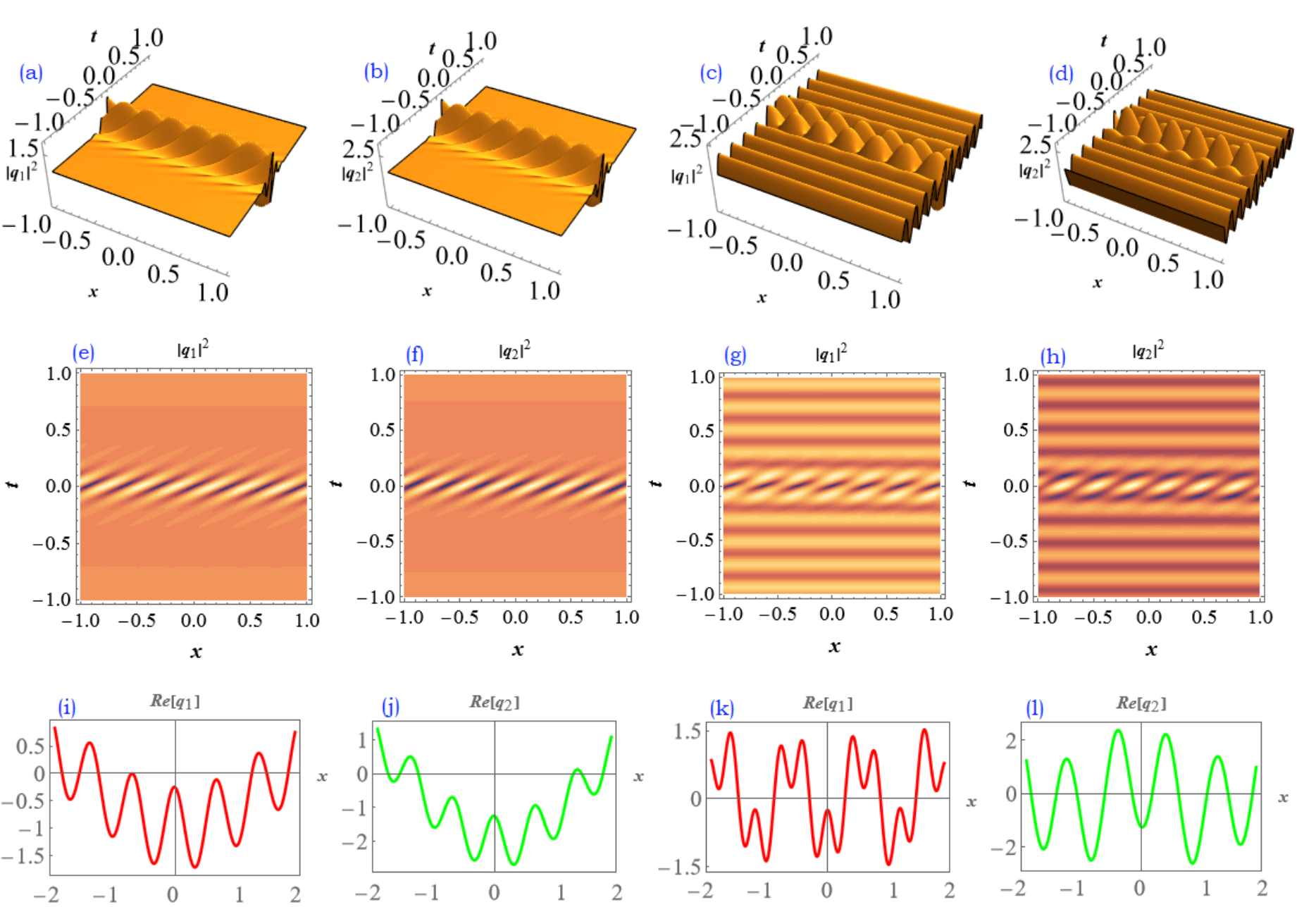}
\caption{\textbf{Breathers With Transient Trap:} The parameter $\sigma(t) = 0.05 t$  while all other parameters are the same as in Figure \ref{Figonebr1}. 
\textbf{First row:} (a, b) Breathers without Rabi coupling ($\Omega = 0$); (c, d) Breathers with Rabi coupling ($\Omega = 2$). \textbf{Middle row:} (e, f) and (g, h)  represent the corresponding contour plots of  the first row  shown by panels (a, b) and (c, d). \textbf{Last row:} (i, j) Real part of the field variables $q_1$ and $q_2$ for $k_L = 0$; (k, l) Real part of the field variables $q_1$ and $q_2$  for $k_L = 8$.}
\label{Figonebr2}
\end{figure}

We have also generated the breathers solutions by choosing the plane wave and zero  as seed solution of the following form
\begin{align}
    q_1[0] &= c e^{i \theta(x,t)}, \\
    q_2[0] &= 0
\end{align}
where $\theta(x,t)= dx+(2 c^2-d^2)t$ and $M= diag\{1,exp(i\theta(x,t)),1)$ and  derive the respective $U_1$ and $V_1$. Then, by analysing the characteristic equation of  $U_1$ and $V_1$ matrices and choosing    $m_1=1, m_2=e^{\tau_1+ik_1},m_3=e^{\tau_2+i k_2},$ where $\tau_{i},k_{i}$ are real numbers, we obtain  breather solution of the following form
\begin{subequations}
\begin{align}
	q_{1}(x,t)&=c_1
	e^{i \theta_i(x,t)} + 6 \left (\zeta _ 1i \right) \frac {\phi _ 1 \phi _ {2}^{\ast} } {\phi _ 1 \phi _ {1}^{\ast} + \phi _ 2 \phi _ {2}^{\ast} + \phi _ 3 \phi _{3}^{\ast}}\\
 q_{2}(x,t)&=c_2
	e^{i \theta_i(x,t)} + 6 \left (\zeta _ 1i \right) \frac {\phi _ 1 \phi _ {2}^{\ast} } {\phi _ 1 \phi _ {1}^{\ast} + \phi _ 2 \phi _ {2}^{\ast} + \phi _ 3 \phi _{3}^{\ast}}
\end{align}\label{Sol:br}
\end{subequations}
where, 
\begin{align}
	\phi_1 &= c e^{\frac{1}{3} i \left(\theta_1(x,t)+\theta_2(x,t)\right)} k_2 m_2+e^{\frac{1}{3} i \left(\theta_1(x,t)+\theta_2(x,t)\right)} k_3
	m_3 \left(\zeta _1+2 i \zeta _1\right) \notag\\
	\phi_{2} &= c e^{-\frac{1}{3} i \left(2 \theta_1(x,t)-\theta_2(x,t)\right)} k_3 m_3+e^{-\frac{1}{3} i \left(2 \theta_1(x,t)-\theta_2(x,t)\right)}
	k_2 m_2 \left(\zeta _1+2 i \zeta _1\right)\notag\\
	\phi_{3} &= e^{-\frac{1}{3} i \left(2 \theta_2(x,t)-\theta_1(x,t)\right)} k_1 m_1\notag\\
 k_1 &=e^{3 i \zeta _1^2 t+i \zeta _1 x}\notag\\
 k_2 &=\exp \left(\zeta _1 x+t \left(2 i \left(c^2+\zeta _1^2\right)+2 \zeta _1 \zeta _1+i \zeta _1^2\right)\right)\notag\\
 k_3 &=\exp \left(\zeta _2 x+t \left(2 i \left(c^2+\zeta _1^2\right)+2 \zeta _1 \zeta _2+i \zeta _2^2\right)\right)\notag
\end{align}
where, $\theta_{i}(x,t)=t \left(2 c^2{}_{i+1}+2 c_i^2-d_i^2\right)+x d_i$, $\zeta_{2}=-\zeta _1+i \left(d-\zeta _1\right), m_2=e^{a_1+i b_1}, m_3= e^{a_2+i b_2}$ with $\zeta_1$ and $\zeta_2$ being  arbitrary complex constants and $a_i,b_i,c_i,d_i$ and $m_1$ are arbitrary real parameters. For a  suitable choice of parameters, the breather solutions without and with trapping potential are displayed in figures. \ref{Figonebr1} and \ref{Figonebr2} respectively. 
\\
\subsection{Breathers without trap}
The breather solution given by Eq.~\eqref{Sol:br} is illustrated in Fig.~\ref{Figonebr1}. Panels (a) and (b ) show the behavior of breathers without Rabi coupling while panels (c) and (d) display breathers with Rabi coupling. The inclusion of Rabi coupling  introduces temporal stripes with double and single mode peaks centred around  $t = 0$ for $q_1$ and $q_2$, as seen in panels (c) and (d). The middle row represents the corresponding contour plots of the first row showing the periodical variation in the amplitude of breathers and the stripes arising by virtue of Rabi coupling. The impact of SO coupling is  illustrated in the last row with the panels (i, j) showing the results for $k_L = 0$ and panels (k, l) for $k_L = 8$. Thus, it is obvious that the  amplitude of the breathers stays constant between a maxima and minima  in a given region of space despite oscillating with time   in the absence of SOC while the addition of SOC  induces rapid fluctuations in the same spatial domain.

\subsection{Breathers with transient trap}
When the trapping potential (as shown in Fig.~\ref{trap}) is switched on, the positioning of the breathers is compressed and tilted, as depicted in panels (a) and (b) of Fig.~\ref{Figonebr2}. The inclusion of Rabi  coupling again  introduces striped bands along the temporal axis but does not significantly impact the double and single mode peaks around $t = 0$ as shown in panels (c) and (d) of Fig.6. The contour plots shown in panels (e),(f),(g) and (h ) confirm the above observation. Comparison of panels (k) and (l)  with (i) and (j)  demonstrates that  one witnesses  increased fluctuations  arising due to the influence of SO coupling in the amplitudes of the field variables $q_1$ and $q_2$.
\subsection{Rogue-Dark-Bright}
\begin{figure}
\centering
\includegraphics[scale=0.40]{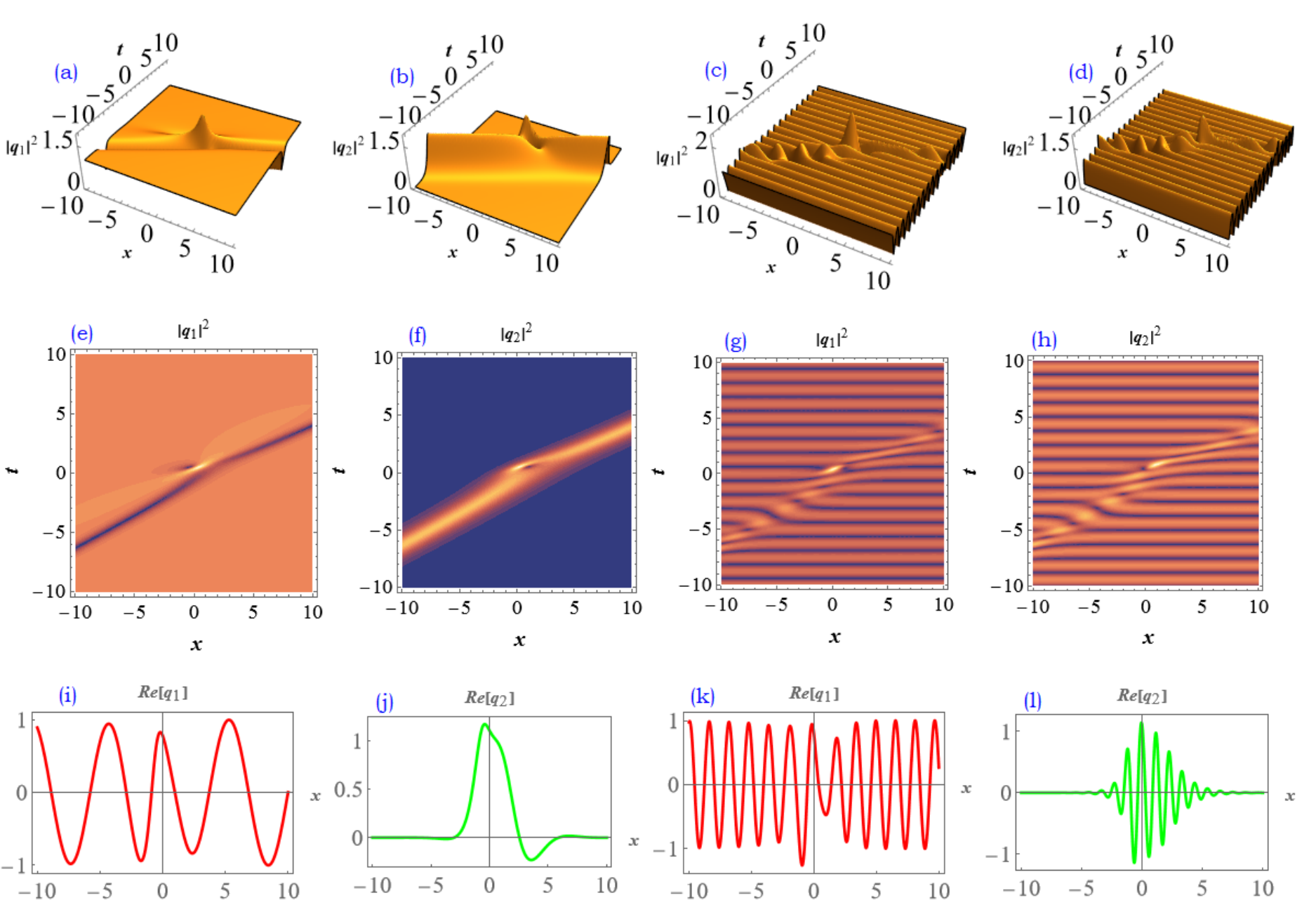}
\caption{\textbf{Mixed Bound States Without Trap:} The parameters  $\sigma(t) = 0$, $a_i = b_i = c = d = 1$. \textbf{First row:} (a, b) Dark-Bright-Rogue waves without  Rabi coupling ($\Omega = 0$), (c, d) Dark-Bright-Rogue waves with Rabi coupling ($\Omega = 2$). \textbf{Middle row:} (e, f) and (g, h) represent the corresponding contour plots of the panels  (a, b) and (c, d), respectively shown in the first row. \textbf{Last row:} (i, j) Real part of the field variables $q_1$ and $q_2$ without SO coupling ($k_L = 0$); (k, l) Real part of the field variables $q_1$ and $q_2$ with SO coupling ($k_L = 8$).}
\label{Figonedbr1}
\end{figure}

\begin{figure}
\centering
\includegraphics[scale=0.40]{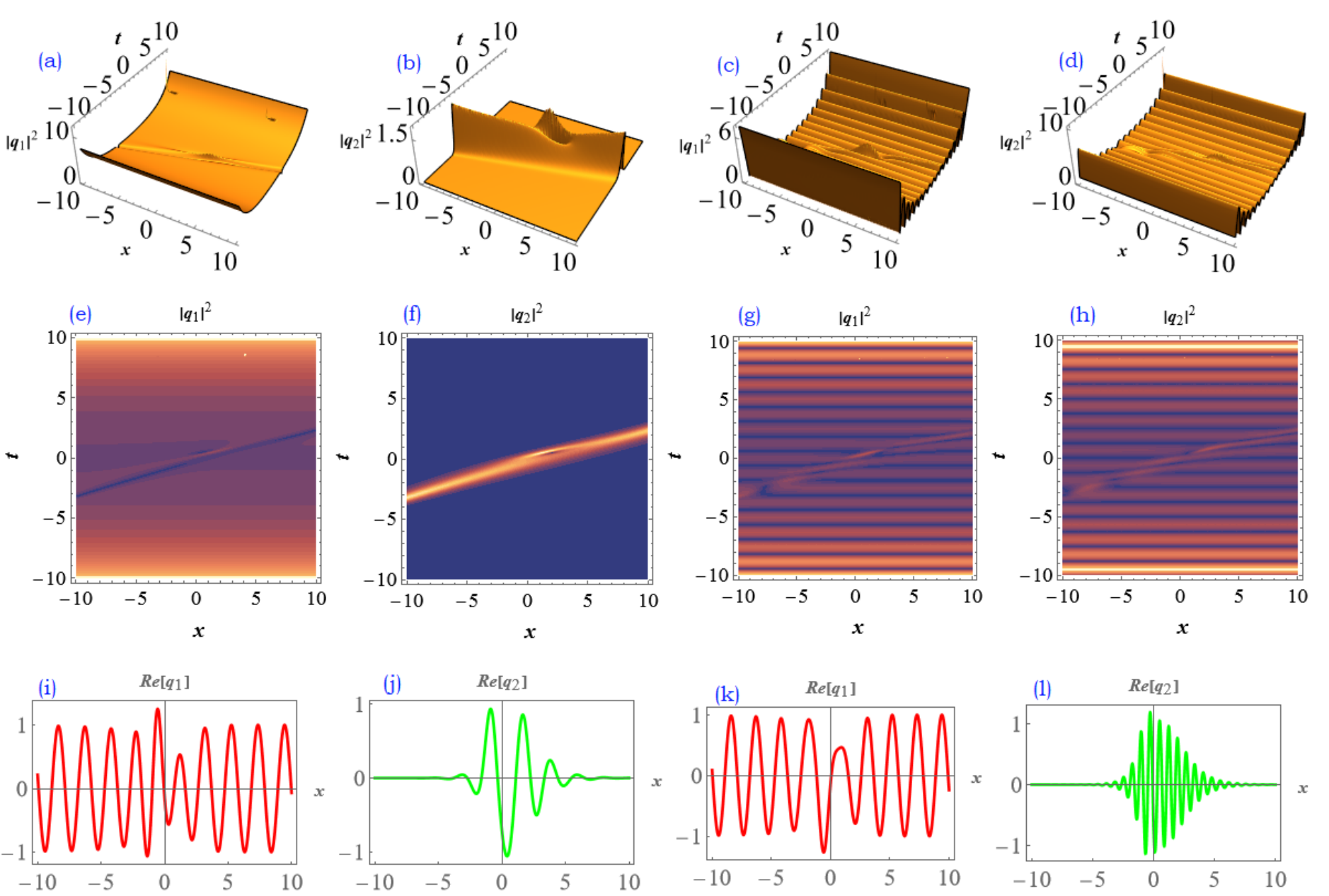}
\caption{\textbf{Mixed Bound States With Transient Trap:} The parameter $\sigma(t) = 0.05 t$ is used, while all other parameters are the same as in Fig.~\ref{Figonedbr1}. \textbf{First row:} (a, b) Dark-Bright-Rogue waves without Rabi coupling ($\Omega = 0$); (c, d) Dark-Bright-Rogue waves with Rabi coupling ($\Omega = 2$). \textbf{Middle row:} (e, f) and (g, h) represent the corresponding contour plots of the  panels (a, b) and (c, d), respectively shown in the first row. \textbf{Last row:} (e, f) Real part of the field variables $q_1$ and $q_2$ without SO coupling ($k_L = 0$) and with SO coupling ($k_L = 8$).}
\label{Figonedbr2}
\end{figure}

One can also superimpose rogue waves with dark and bright soliton solutions by choosing the same seed as discussed  earlier in the construction of breathers  and continue the Darboux transformation approach until we obtain  $U_1$ and $V_1$ matrices. By analyzing the charateristic equation of the matrices and its roots and   choosing  multiple root options instead of a single root along with the  the parametric choice $m_1=a_1+b_1 i, m_2=1,m_3=a_3+b_3 i$, we are able to construct the amalgamated  Rogue-dark-bright solution which are termed as  "mixed bound states"  of the following form:
\begin{subequations}
\begin{align}
	q_{1}(x,t) &= c e^{i \theta(x,t) } \left[1-\frac{4 e^{-\text{$\eta $1R}} \left(\alpha ^2-\alpha +\beta ^2+i \beta
		\right)}{e^{-\text{$\eta $1R}} \left[2 \left(\alpha ^2+\beta ^2\right)-2 \alpha +1\right]+e^{2 \text{$\eta $1R}}
		\left[a_3^2+b_3^2\right]}\right]\\
		q_{2}(x,t) &= -\frac{4 c \left[(\alpha -1)\; a_3+i \left(a_3 \beta +(\alpha -1)\; b_3\right)-\beta\;  b_3\right] e^{\frac{\text{$\eta
						$1R}}{2}+i \left[t \left(3 c^2-d^2\right)+d x\right]}}{e^{-\text{$\eta $1R}} \left(2 \left(\alpha ^2+\beta
			^2\right)-2 \alpha +1\right)+e^{2 \text{$\eta $1R}} \left(a_3^2+b_3^2\right)}
\end{align}\label{ro-db}
\end{subequations}
where,
$\theta(x,t) = d x + \left(2 c^2-d^2\right) t$, $\mu = \frac{1}{2} \log \left(\frac{a_3^2+b_3^2}{2 c^2}\right)$, $\eta_{1R} = \frac{2}{3} c (x-2 d t)$, $\alpha=a_1 c+c (x-2 d t)$ and $\beta= b_1 c-2 c^2 t$ with $a_i, b_i, c$ and $d$ being arbitrary real parameters. Again, for a  suitable choice of parameters, the mixed bound states with rogue waves are displayed in the figure. \ref{Figonedbr1} and \ref{Figonedbr2} without and with transient trap respectively.

\subsection{Mixed Bound States Without  Transient Trap}
The collision of rogue waves with dark and bright solitons, as given by Eq.~\eqref{ro-db} in the absence of harmonic  trap  without Rabi coupling is shown in panels (a) and (b) of Fig.~\ref{Figonedbr1} . Introduction of  Rabi coupling in panels  results in striped bands along the temporal axis, as seen in panels (c, d). The contour plots in the middle row shown by panels (g) and (h) reconfirm the impact of Rabi coupling resulting in the formation of striped bands in the temporal direction while one does not observe any such stripe bands in the panels (e) and (f). Comparison of  panels (i)and (j) with (k) and (l) highlights the impact of  SO coupling which introduces   rapid fluctuations in the amplitude of the real part of the  rogue-dark-bright solitons. 

\subsection{ Mixed Bound States with Transient Trap} Introduction  of  the time dependent trap compresses the amplitude of mixed bound state  shown in panels (a) and (b) of Fig.~\ref{Figonedbr2}  which is further confirmed by the contour plots shown in panels (e) and (f). The inclusion of Rabi  coupling  creates striped bands again shown by panels (g) and (h). In addition, when the mixed bound states cross over to the expulsive domain, their amplitude overshoots the upper bound   as shown in panels (c) and (d) of Fig.~\ref{Figonedbr2}.  As observed  earlier,  the rapid oscillations due to SO coupling are more pronounced in panels (k) and  (l)  as compared to panels (i) and  (j).

\subsection{Dark-Bright-Solitons without and with transient trap}
\begin{figure}
\centering
\includegraphics[scale=0.40]{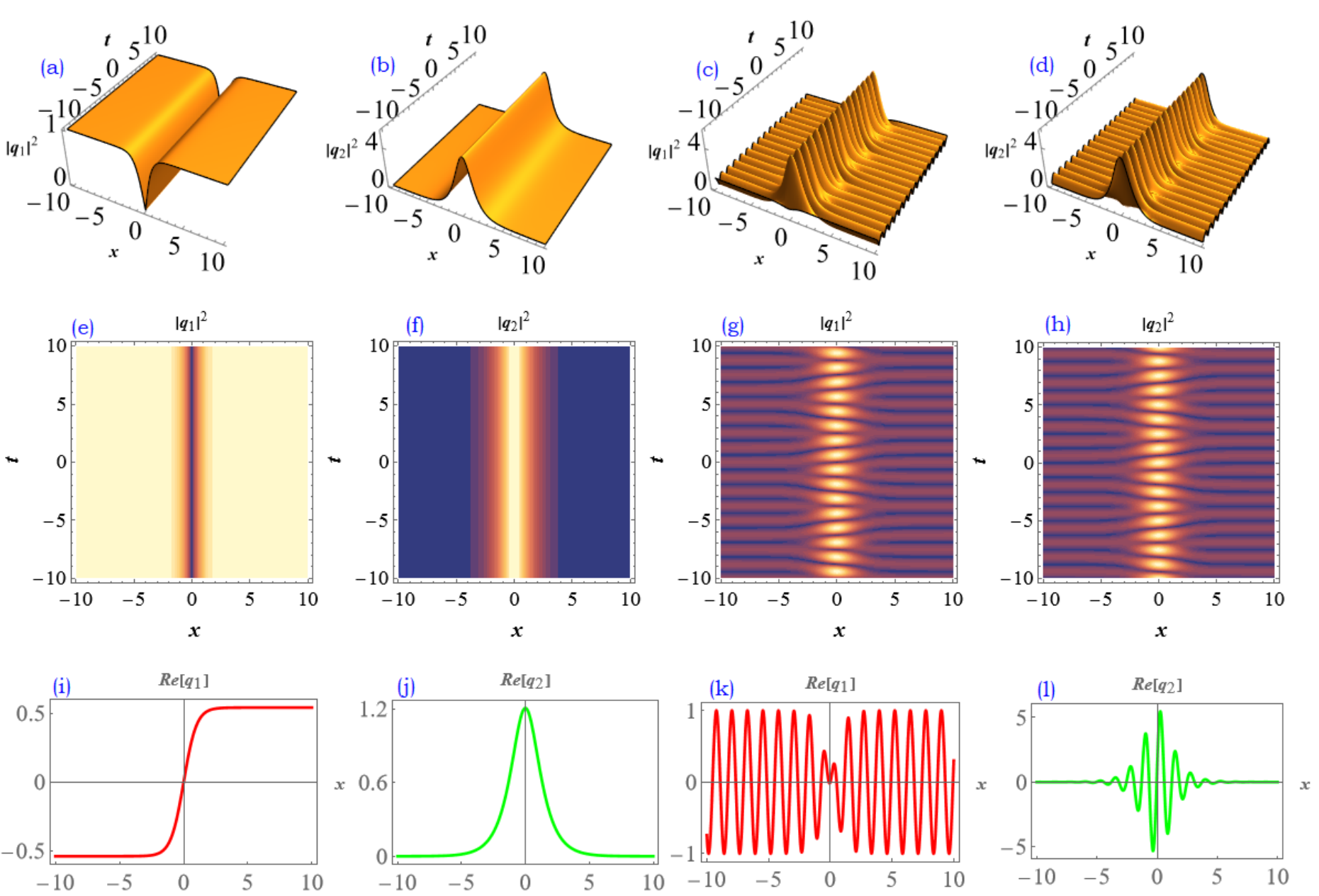}
\caption{Dark-Bright solitons without trap for  $\sigma(t)=0, a_3=b_3=c=1,$ and $d=0$ \textbf{First row} (a,b) Dark-Bright-without Rabi coupling ($\Omega = 0$) and \textbf{c,d}  Dark-Bright- Rogue waves with Rabi coupling $\Omega = 2$. \textbf{Middle row} (e,f) represent the contour plots of (a,b) and (g,h) contour plots (c,d) respectively. \textbf{Last row} (i,j) and (k,l) show the real part of the field variables $q_1$ and $q_2$ without and with  SO coupling for $k_L = 8$ respectively} \label{Figonedb1}
\end{figure}

\begin{figure}
\centering
\includegraphics[scale=0.40]{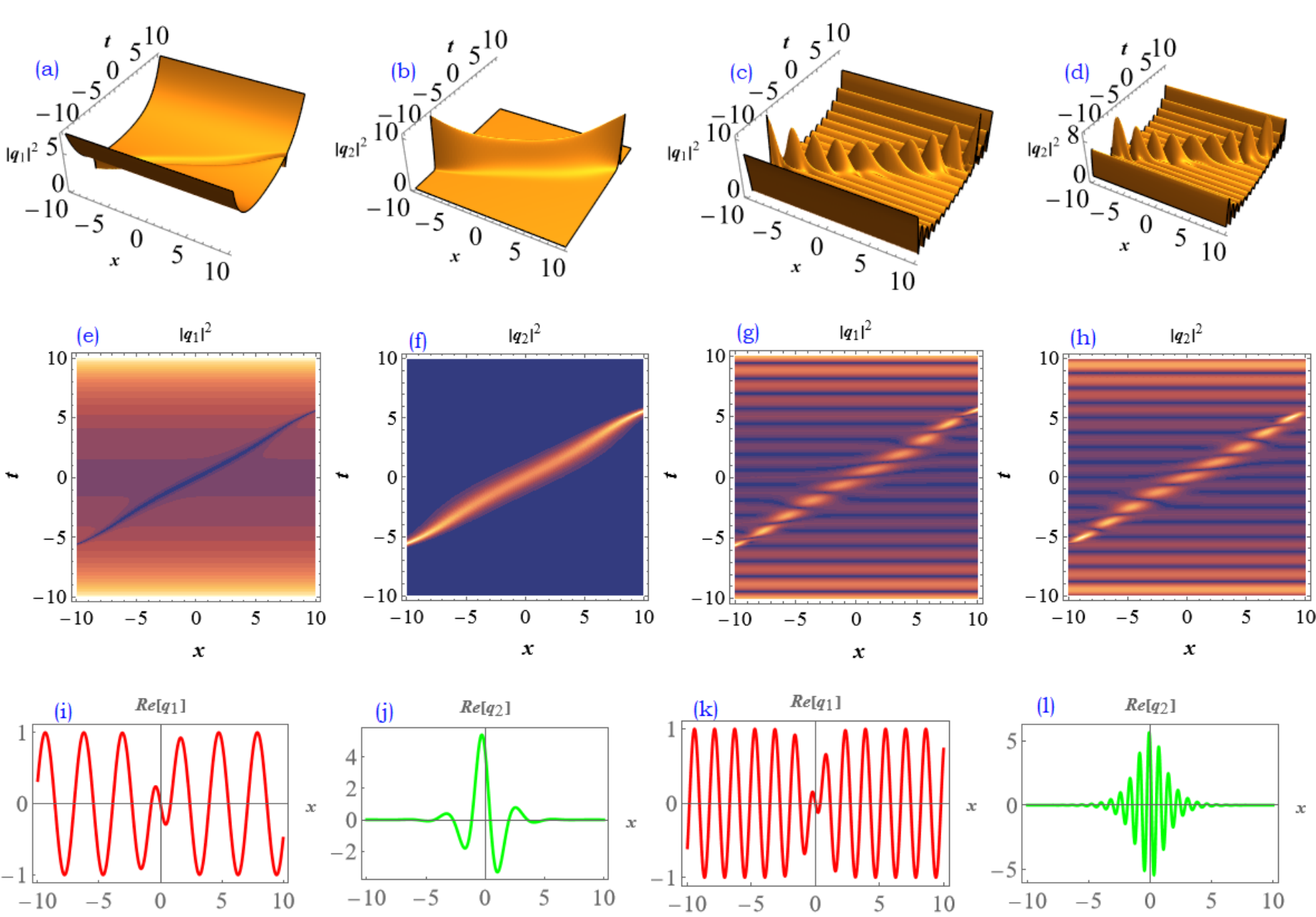}
\caption{Dark-Bright solitons With transient trap for $\sigma(t)=0.05 t$ with  all other parameters being the  same as in  fig.\ref{Figonedb1}. \textbf{First row} (a,b) and (c,d) Dark-Bright solitons without and with Rabi coupling for ($\Omega$ =0) and ($\Omega = 2$) respectively. \textbf{Middle row} (e,f) and  (g,h) represent the corresponding contour plots of  the first row . \textbf{Last row} (i,j) and (k,l) illustrate the real part of the field variables $q_1$ and $q_2$ without and with  SO coupling for $k_L = 8$}\label{Figonedb2}
\end{figure}

The classical dark- bright soliton solution is generated for  the  following  choice of parameters  $m_1=1, m_2=0,m_3=a_3+b_3 i$ to obtain
\begin{subequations}
\begin{align}
	q_{1}(x,t) &= c e^{i \theta(x,t) } \tanh \left[\frac{3 \text{$\eta_{1R} $}}{2}+\mu \right] \\
	q_{2}(x,t) &= -4 c^2 \left(a_3+i b_3\right) \text{sech}\left[\frac{3 \text{$\eta_{1R} $}}{2}+\mu \right] e^{i \left[t \left(3
		c^2-d^2\right)+d x\right]}
\end{align}\label{db_sol}
\end{subequations}
where the parameters $\theta(x,t)$, $\mu$ and $\eta_{1R}$ are the  same as in the previous section.   The profiles of dark and bright solitons without and with trapping potential are  displayed in  figures. \ref{Figonedb1} and \ref{Figonedb2} respectively.
\\

The conventional trapless dark and bright solitons without  Rabi coupling, are shown in panels (a) and (b) of  Fig.~\ref{Figonedb1} and is complimented by the contour plots shown by panels (e) and (f) .  We observe that the inclusion of Rabi coupling  introduces stripes  along the temporal axis in the density profile of both dark and bright solitons as witnessed in panels (c) and (d) which is again confirmed by the contour plots shown by panels (g) and (h) . In addition, we also see the flipping of dark solitons (shown by panel (a)) to attain positive density profile as shown in panel (c) quite similar to a bright soliton and this occurs by virtue of Rabi coupling.  From the panels (i),(j),(k) and (l) shown in the last row, we notice that the addition of SO coupling introduces rapid fluctuations in the dark solitons while the fluctuations are centred around the origin in the bright solitons.

When we switch on the transient trap (shown in Fig.~\ref{trap} ), one witnesses a 45$^{\circ}$ shift in the trajectory of both dark and bright solitons shown  in panels (a) and (b) which are further endorsed by the contour plots shown in panels (e)and (f) in addition to the stripes due to Rabi coupling shown in panels (c) and (d) which are in confirmity with the panels (g) and (h).  In addition, we also see that the width of the solitons widen in the confining trap while they shrink in the expulsive domain.Besides, we also observe that when the  transient trap is switched on, it introduces oscillations in the real part of the order parameters $q_1$ and $q_2$ (shown in panels (i) and (j)) which are further enhanced by the addition of SO coupling shown in panels (k) and (l).  It can also be noticed that the amplitude of real part of $q_1$ around origin almost becomes zero for dark solitons while that of $q_2$ becomes maximum around  origin.

\subsection{Second order Rogue wave}
\begin{figure}
 \centering
 \includegraphics[scale=0.90]{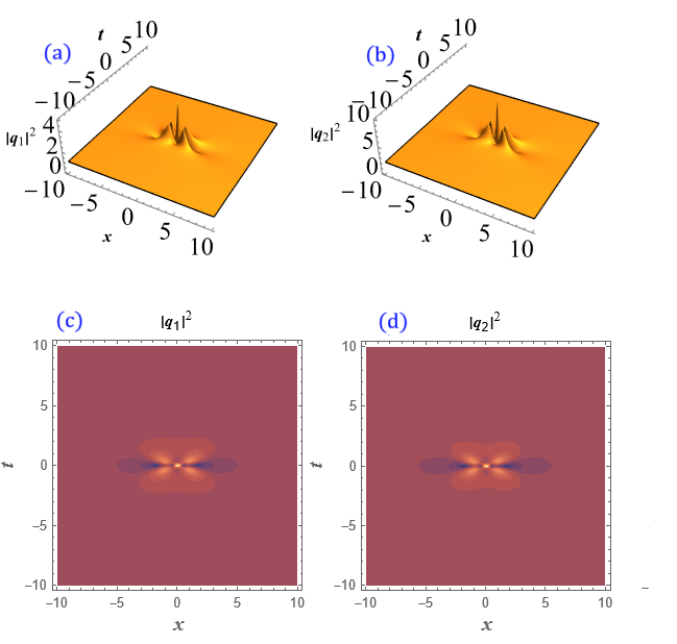}
\caption{ Panels (a) and (b) represent a second order rogue wave  for $\sigma(t)=0,c_1=1,c_2=2,m_1=0.1,m_2=0.5$ with a primary crest  at the centre  surrounded by four secondary crests (c), (d) contour plots of  (a,b) with a  four petal  structure.}\label{2rog}
\end{figure}
Darboux transformation approach can be extended to generate multi rogue waves. For example, the  second order rogue waves are  of the following form:
\begin{subequations}
\begin{align}
	q_{1}(x,t) &= c_1 \; e^{2 i t} \left[1+\frac{G_{2}(x,t)+i\; H_{2}(x,t)\; t}{d_{2}(x,t)}\right] \\
	q_{2}(x,t) &= c_2 \; e^{2 i t} \left[1+\frac{G_{2}(x,t)+i\; H_{2}(x,t)\; t}{d_{2}(x,t)}\right]
\end{align}\label{tworog}
\end{subequations}
\begin{align}
    G_{2}(x,t) &= m_2^2 \left(-7680 t^4-2304 t^2 x^2-1728 t^2-96 x^4-144 x^2+72 x+18\right)-384 m_1 m_2 x \notag\\
    H_{2}(x,t) &= m_2^2 \left(-6144 t^4-3072 t^2 x^2-768 t^2-384 x^4-576 x^2+288 x+360\right)-1536 m_1 m_2 x \notag\\
    d_{2}(x,t) &= 128 m_1^2+m_1 m_2 \left(1536 t^2 x-128 x^3+96 x-48\right)+\notag\\
    &m_2^2 \left(2048 t^6+1536 t^4
   x^2+3456 t^4+384 t^2 x^4-576 t^2 x^2-288 t^2 x+792 t^2+32 x^6+24 x^4+24 x^3+54 x^2-18 x+9\right)\notag
\end{align}
For a  suitable choice of parameters, second order rogue waves with a four petal structure  are  shown in panels (c) and (d) of fig.\ref{2rog} while we observe a primary crest with a peak intensity at the centre surrounded by four secondary crests shown in  panels (a) and (b). 
\section{Conclusion}
In this paper, we have investigated the spin orbit -Rabi coupled   condensates described  by the two coupled Gross-Pitaevskii(GP) equation  based on the interplay between SOC, Rabi coupling,transient trap frequency and time dependent scattering length. Employing Darboux transformation approach with nontrivial seed solution, we generate rogue waves, breathers, mixed rogue-dark-bright and classical dark-bright  solutions. While the addition of spin orbit coupling introduces rapid oscillations in the amplitude of the condensates, Rabi coupling introduces  stripes in the temporal direction of the condensates. We have also extensively brought out the interplay between SOC, Rabi coupling, trap frequency and scattering length to showcase the rich dynamics of the condensates. The fact that the above wave phenomena like rogue waves, breathers etc,. have not yet been generated analytically adds to the physical significance of the results. In addition, the interplay between SOC, Rabi coupling, trap frequency and scattering length will certainly turn out to be an ideal testbed  for experiments. We do really hope the identification of the above nonlinear excitations  will definitely motivate researchers to unearth them experimentally.  

\section{Declaration of Competing Interest}
The authors declare that they  have no  competing financial interests or personal relationships that could have appeared to influence the work reported in this paper.
	
\section{Acknowledgments}
Authors are indebted to the Referees for their comments/suggestions which have contributed to the improvement in the presentation of the results. PSV wishes to express his deepest gratitude to the Principal and the Management of PSG College of Arts and Science for their moral support and encouragement throughout the tenure of this project. RR wishes to thank DST-CURIE for financial assistance (DST/CURIE -PG/2022/54)  and DST-CRG (CRG/2023/008153) for financial support.

\section{Data availability statement}
Data sets generated to represent the the current study are available within the article.

\end{document}